\documentclass[ reprint,  amsmath,amssymb,  aps, ]{revtex4-1}%
\usepackage{graphicx}
\usepackage{dcolumn}
\usepackage{bm}
\usepackage{amsmath}%
\usepackage{amsfonts}%
\usepackage{amssymb}

\begin{document}

\title{ }%

\title
{Semi-analytic theory of self-similar optical propagation and mode-locking
using a shape-adaptive model pulse}
\author{Christian Jirauschek}
\email[]{jirauschek@mytum.de}
\affiliation{Institute for Nanoelectronics, Technische Universit{\"a}%
t M{\"u}nchen, Arcisstra{\ss}e 21, D-80333 Munich, Germany}
\author{F. {\"O}mer Ilday}
\affiliation{Department of Physics, Bilkent University, 06800, Ankara, Turkey}
\date{\today, published as Phys. Rev. A 83, 063809 (2011)}
\begin{abstract}
A semi-analytic theory for the pulse dynamics in similariton amplifiers and
lasers is presented, based on a model pulse with adaptive shape. By changing a
single parameter, this test function can be continuously tweaked between a pure
Gaussian and a pure parabolic profile, and can even represent sech-like
pulses, the shape of a soliton. This approach allows us to describe the pulse
evolution in the self-similar and other regimes of optical propagation.
Employing the method of moments, the evolution equations for the
characteristic pulse parameters are derived from the governing nonlinear Schr{\"o}%
dinger/Ginzburg-Landau equation. Due to its greatly reduced complexity, this description allows for extensive parameter optimization, and can aid intuitive understanding of
the dynamics. As an application of this approach, we model a soliton-similariton laser and validate the results against numerical simulations. This constitutes a semi-analytic model of the soliton-similariton laser. Due to the versatility of the
model pulse, it can also prove useful in other application areas.
\end{abstract}
\maketitle

\section{\label{sec:level1}Introduction}

Self-similarity is a recurring theme in strongly nonlinear systems. Its
observation can be particularly informative as it implies an underlying
symmetry, which can be exploited mathematically through symmetry reduction
techniques~\cite{symmetry-reduction}. In nonlinear optics, self-similarity
emerges in the formation of Cantor-set fractals in materials that support
spatial solitons \cite{Cantor}, the self-collapse of beams at high powers
\cite{mol03}, and in the propagation of ultrafast pulses of light in optical
fiber amplifiers in the presence of strong Kerr
nonlinearity~\cite{parabolic-amp1,parabolic-amp2}. In recent years, it was
reported that self-similar propagation of short pulses in laser resonators is
possible~\cite{SS_PRL,SS_TiSa}. These pulses have a nearly parabolic intensity
profile and evolve self-similarly within the nonlinear segments of the laser
cavity. Fiber lasers supporting self-similarly evolving pulses is now
recognized as new regime of pulse formation in the cavity of an ultrafast
laser. This method is differentiated from the well-known
solitary~\cite{soliton}, stretched-pulse (dispersion-managed)~\cite{SP} and
all-normal-dispersion~\cite{all-normal} solutions to the Haus Master
equation~\cite{SS_Haus_theory}. There are interesting similarities as well as
important differences between these regimes. From a practical point of view,
the demonstration of the similariton laser has led to the development of fiber
lasers with significantly higher pulse energies~\cite{10nJSS}. These fiber
lasers are being studied by many groups \cite{nie05,rue06,an07,ren10},
motivated by the various applications ultrafast lasers have in diverse areas
of physics, from optical frequency metrology and material processing to
next-generation accelerators. More recently, a new mode-locking regime, the
soliton-similariton laser was reported, in which the pulse evolution is in the
form of periodic alteration between soliton and similariton
evolution~\cite{SS_NP}. One aspect of this regime is that the evolution is
strongly nonlinear at every point in the laser cavity. The possibilities and
limitations of this regime are largely in need of exploration, for which
theoretical modeling is crucial. For all of these reasons, there is much
desire to understand the physics of amplifier similaritons and self-similar
lasers better.

Numerical simulations provide good agreement with experiments
\cite{SS_PRL,rue06,SS_NP}. However, they are computationally expensive,
rendering extended explorations of the parameter space impractical. Moreover,
a theoretical description can aid intuitive understanding of the dynamics of
self-similar evolution in optical amplifiers and lasers. Exact self-similar
solutions have been derived for the optical pulse propagation in fibers with
and without gain~\cite{parabolic-amp1,parabolic-amp2,and93}. However, the
pulse shape evolves during propagation, and the self-similar parabolic pulse
profile is only asymptotically reached. Thus, several approaches have been
explored to derive a simplified description which still captures the rich
pulse dynamics in such systems. Based on various analytical methods, the pulse
formation, pulse stability and energy scalability of similariton and other
high-energy fiber lasers has been studied \cite{bal08b,bel07,kal10}. Also
semi-analytic approaches, widely used in optics to investigate pulse
propagation, have been employed. They aim to extract evolution equations for
characteristic pulse parameters, reducing the partial differential equation
for pulse propagation to a coupled set of ordinary differential equations.
Such approaches are typically based on the method of moments (MOM) or a
variational formalism, which have both been used to investigate the evolution
of the pulse energy and the temporal and spectral pulse width in the strongly
nonlinear regime \cite{bur07,ant07,bal10}. Such studies typically rely on
fixed pulse shapes such as Gaussian or sech pulses, yielding reasonable
estimates for the pulse energy and duration, but no pulse shape information at
all. An exception can be found in \cite{and07}, where an adaptive
super-Gaussian test function was used to investigate changes of the pulse
profile during propagation.

Here, we report on a semi-analytic theory for the pulse dynamics in
similariton amplifiers and lasers including the soliton-similariton laser,
based on a novel model pulse with adaptive shape. The key in this formulation
is our ansatz function that can describe any pulse shape from a pure Gaussian
to a pure parabolic profile, even including $\mathrm{sech}$-like pulses (i.e.,
with $\mathrm{sech}^{2}$ intensity profile), the shape of a soliton. The pulse
profile is tweaked by a single parameter, which is complemented by an
additional degree of freedom for the pulse phase. This allows us to represent
various pulse profiles as well as complex spectral shapes. Thus, our
theoretical treatment appears to be capable of describing not only the
self-similar but the other regimes as well, opening the way to a simple
unified theoretical approach.

Employing the method of moments \cite{MoM,tso06}, the partial differential
equation governing the pulse propagation is reduced to a finite set of coupled
ordinary differential equations, which are much easier to analyze. In
addition, the coefficients of the equations are helpful in forming an
intuitive understanding of the dynamics by exposing the relative importance of
the various effects. Through investigation of these equations one gains access
to valuable information about the pulse dynamics, e.g., of how exactly the
various effects on the pulse are paired to balance each other to satisfy the
periodic boundary conditions imposed by the laser resonator. Such information
is extremely difficult, if not impossible, to obtain by repeated numerical
solutions of the full governing equation. Our approach is validated against
numerical results for single-pass propagation and for the steady state
dynamics of a soliton-similariton laser.

\section{Test Pulse and Evolution Equations}

For propagation through a dispersive Kerr medium with a parabolic gain and
instantaneously saturable absorption, the evolution of the pulse envelope
$u(z,t)$ is described by a generalized nonlinear Schr\"{o}dinger (or complex
Ginzburg-Landau) equation of the form \cite{agr89}
\begin{equation}
\mathrm{i}\partial_{z}u-D\partial_{t}^{2}u+\gamma\left|  u\right|
^{2}u=\mathrm{i}\left(  g+g_{\omega}\partial_{t}^{2}+r\left|  u\right|
^{2}\right)  u. \label{nse}%
\end{equation}
Here, $z$ and $t\ $are the propagation coordinate and the retarded time,
respectively. $D$ is the second order dispersion coefficient, and $\gamma$ is
the cubic nonlinearity parameter. The dissipative processes are characterized
by the central gain value $g$ and spectral gain parameter $g_{\omega}$ as well
as the saturable absorption coefficient $r$. Generally, $D$, $\gamma$, $g$,
$g_{\omega}$, and $r$ are $z$ dependent, since an optical system such as a
fiber laser consists of a sequence of different segments. Additionally, the
parameter values can vary even within a segment, for example $g$ if gain
saturation is considered.

\subsection{Test Pulse}

For linear systems, $\gamma=r=0$, the complex Gaussian%
\begin{equation}
u\left(  z,\tau\right)  =A\sqrt{p_{1}(\tau)}\exp\left(  \mathrm{i}\beta
\tau^{2}+\mathrm{i}\phi\right)  \label{gauss}%
\end{equation}
with $p_{1}(\tau)=\exp\left(  -\tau^{2}\right)  $ is an exact solution to
Eq.\thinspace(\ref{nse}), where $\tau=t/T$ denotes the normalized time, and
$T(z)$, $A(z)$, $\phi(z)$ and $\beta(z)$ are the pulse duration, amplitude,
phase and linear chirp parameter, respectively. Thus, for moderate
nonlinearity, the Gaussian ansatz is still a good description of the
steady-state pulse shape in a laser cavity \cite{che99,jir03,jir06}. In
contrast, in the strongly nonlinear limit, the pulse is approximately
described by a self-similar pulse with a parabolic intensity profile. However,
an exactly parabolic pulse is an idealization and in practice the pulse shape
is parabolic around the center, where most of the energy resides, but with a
super-Gaussian fall-off in the wings~\cite{SS_PRL,and93}. Naturally, in the
intermediate regime, the pulse shape combines features of a Gaussian pulse and
a self-similar pulse. To reflect these properties, we have previously
introduced a function of the type
\begin{equation}
p_{n}\left(  \tau\right)  =\exp\left(  -\sum_{k=1}^{n}\tau^{2k}/k\right)
=1-\tau^{2}+\mathcal{O}\left(  \tau^{2n+2}\right)  \label{pn}%
\end{equation}
to describe the pulse profile, which represents a Gaussian for $n=1$ and a
parabolic profile for $n\rightarrow\infty$ \cite{jir06b}. Here, the pulse
duration $T$ represents the Gaussian pulse width for $n=1$ and half the total
pulse width of a similariton for $n\rightarrow\infty$. This ansatz has been
shown to be useful for the description of similariton lasers and trapped
Bose-Einstein condensates \cite{jir06b,kec07}.

A disadvantage of Eq.\thinspace(\ref{pn}) is that the pulse shape cannot be
adapted continuously, but only in discrete steps. Using the Gauss
hypergeometric function $_{2}$\textrm{F}$_{1}$ for which efficient numerical
evaluation routines exist \cite{nr}, Eq.\thinspace(\ref{pn}) can be expressed
in closed form as
\begin{equation}
p_{n}\left(  \tau\right)  =\left(  1-\tau^{2}\right)  \exp\left\{
\frac{\left|  \tau\right|  ^{2n}}{n}\left[  _{2}\mathrm{F}_{1}\left(
1,n;1+n;\tau^{2}\right)  -1\right]  \right\}  , \label{pn2}%
\end{equation}
see also Appendix \ref{Aa}. In Eq.\thinspace(\ref{pn2}), $n$ is not restricted
to integers, providing much more flexibility for describing different pulse
shapes. For example, $\mathrm{sech}^{2}$-like intensity profiles,
corresponding to a fundamental optical soliton, are very well represented by
$n\approx0.5$. Moreover, rather than a priory fixing $n$ to a certain value,
we allow $n=n\left(  z\right)  $ to evolve during pulse propagation,
describing the position dependent intensity profile together with the
parameters $A\left(  z\right)  $ and $T\left(  z\right)  $. Along with
$n\left(  z\right)  $, the third order chirp parameter $\alpha\left(
z\right)  $ is introduced as a further degree of freedom for the pulse phase
in addition to $\beta\left(  z\right)  $ and $\phi\left(  z\right)  $, to
avoid mathematical problems with the evolution equations for the pulse
parameters \cite{comment2}. The resulting ansatz for the envelope is given by%
\begin{equation}
u\left(  z,\tau\right)  =A\sqrt{p_{n}(\tau)}\exp\left(  \mathrm{i}\beta
\tau^{2}+\mathrm{i}\alpha\tau^{4}+\mathrm{i}\phi\right)  . \label{ansatz}%
\end{equation}
Naturally, Eq.\thinspace(\ref{pn2}) is not the only function which is able to
interpolate continuously between a parabolic and a Gaussian shape. In
particular, the so-called q-Gaussian function \cite{boz97} has been used in
various contexts, e.g., for the description of trapped Bose-Einstein
condensates \cite{nic08}. While the q-Gaussian has a somewhat simpler
analytical form, it is non-zero only on a finite interval (except for the
limiting case of a Gaussian), which is unphysical for the applications
considered in this paper. Additionally, our ansatz has the distinct advantage
that it can also represent a $\mathrm{sech}^{2}$ profile to a very good
approximation, which is essential for a versatile description of nonlinear
optical propagation.

\subsection{Evolution Equations}

The generalized nonlinear Schr\"{o}dinger equation Eq.\thinspace(\ref{nse})
can be approximately solved by extracting evolution equations for the
parameters of the model pulse Eq.\thinspace(\ref{ansatz}). Here we use the
method of moments (MoM) \cite{MoM}; the derivation can be found in Appendix
\ref{Ab}. The resulting equations of motion are%
\begin{align}
n^{\prime}  &  =\left\{  2rA^{2}\left(  \frac{\mu_{4}}{\varepsilon_{4}}%
-2\frac{\mu_{2}}{\varepsilon_{2}}+\frac{\mu_{0}}{\varepsilon_{0}}\right)
+32\alpha DT^{-2}\left(  \frac{\varepsilon_{4}}{\varepsilon_{2}}%
-\frac{\varepsilon_{6}}{\varepsilon_{4}}\right)  \right. \nonumber\\
&  -g_{\omega}T^{-2}\left[  \frac{1}{2}\frac{\eta_{0}}{\varepsilon_{0}}%
-\frac{\eta_{2}}{\varepsilon_{2}}+4\frac{\varepsilon_{0}}{\varepsilon_{2}%
}\mathbb{+}\frac{1}{2}\frac{\eta_{4}}{\varepsilon_{4}}-12\frac{\varepsilon
_{2}}{\varepsilon_{4}}\right. \nonumber\\
&  +8\beta^{2}\left(  \frac{\varepsilon_{2}}{\varepsilon_{0}}-2\frac
{\varepsilon_{4}}{\varepsilon_{2}}+\frac{\varepsilon_{6}}{\varepsilon_{4}%
}\right)  +32\alpha^{2}\left(  \frac{\varepsilon_{6}}{\varepsilon_{0}}%
-2\frac{\varepsilon_{8}}{\varepsilon_{2}}\mathbb{+}\frac{\varepsilon_{10}%
}{\varepsilon_{4}}\right) \nonumber\\
&  \left.  \left.  +32\beta\alpha\left(  \frac{\varepsilon_{4}}{\varepsilon
_{0}}-2\frac{\varepsilon_{6}}{\varepsilon_{2}}+\frac{\varepsilon_{8}%
}{\varepsilon_{4}}\right)  \right]  \right\} \nonumber\\
&  \left/  \left(  \frac{\partial_{n}\varepsilon_{0}}{\varepsilon_{0}}%
-2\frac{\partial_{n}\varepsilon_{2}}{\varepsilon_{2}}+\frac{\partial
_{n}\varepsilon_{4}}{\varepsilon_{4}}\right)  \right.  , \label{ext_n}%
\end{align}%

\begin{align}
\frac{T^{\prime}}{T}  &  =-4DT^{-2}\left(  \beta+2\alpha\frac{\varepsilon_{4}%
}{\varepsilon_{2}}\right)  +rA^{2}\left(  \frac{\mu_{2}}{\varepsilon_{2}%
}-\frac{\mu_{0}}{\varepsilon_{0}}\right) \nonumber\\
&  +g_{\omega}T^{-2}\left[  \frac{1}{4}\frac{\eta_{0}}{\varepsilon_{0}}%
-\frac{1}{4}\frac{\eta_{2}}{\varepsilon_{2}}+\frac{\varepsilon_{0}%
}{\varepsilon_{2}}+4\beta^{2}\left(  \frac{\varepsilon_{2}}{\varepsilon_{0}%
}-\frac{\varepsilon_{4}}{\varepsilon_{2}}\right)  \right. \nonumber\\
&  \left.  +16\alpha^{2}\left(  \frac{\varepsilon_{6}}{\varepsilon_{0}}%
-\frac{\varepsilon_{8}}{\varepsilon_{2}}\right)  +16\beta\alpha\left(
\frac{\varepsilon_{4}}{\varepsilon_{0}}-\frac{\varepsilon_{6}}{\varepsilon
_{2}}\right)  \right] \nonumber\\
&  +\frac{1}{2}n^{\prime}\left(  \frac{\partial_{n}\varepsilon_{0}%
}{\varepsilon_{0}}-\frac{\partial_{n}\varepsilon_{2}}{\varepsilon_{2}}\right)
, \label{ext_T}%
\end{align}%

\begin{align}
\frac{A^{\prime}}{A}  &  =2DT^{-2}\left(  \beta+2\alpha\frac{\varepsilon_{4}%
}{\varepsilon_{2}}\right)  +g+\frac{1}{2}rA^{2}\left(  3\frac{\mu_{0}%
}{\varepsilon_{0}}-\frac{\mu_{2}}{\varepsilon_{2}}\right) \nonumber\\
&  +g_{\omega}T^{-2}\left[  -\frac{3}{8}\frac{\eta_{0}}{\varepsilon_{0}}%
+\frac{1}{8}\frac{\eta_{2}}{\varepsilon_{2}}-\frac{1}{2}\frac{\varepsilon_{0}%
}{\varepsilon_{2}}+2\beta^{2}\left(  \frac{\varepsilon_{4}}{\varepsilon_{2}%
}-3\frac{\varepsilon_{2}}{\varepsilon_{0}}\right)  \right. \nonumber\\
&  \left.  +8\alpha^{2}\left(  \frac{\varepsilon_{8}}{\varepsilon_{2}}%
-3\frac{\varepsilon_{6}}{\varepsilon_{0}}\right)  +8\beta\alpha\left(
\frac{\varepsilon_{6}}{\varepsilon_{2}}-3\frac{\varepsilon_{4}}{\varepsilon
_{0}}\right)  \right] \nonumber\\
&  +\frac{1}{4}n^{\prime}\left(  \frac{\partial_{n}\varepsilon_{2}%
}{\varepsilon_{2}}-3\frac{\partial_{n}\varepsilon_{0}}{\varepsilon_{0}%
}\right)  , \label{ext_A}%
\end{align}%
\begin{align}
\alpha^{\prime}  &  =4\frac{T^{\prime}}{T}\alpha+\bigg\{2g\alpha
\varepsilon_{6}+2rA^{2}\alpha\mu_{6}\nonumber\\
&  +\frac{1}{2}g_{\omega}T^{-2}\bigg[\beta\left(  9\varepsilon_{2}%
-\frac{\varepsilon_{0}\varepsilon_{4}}{\varepsilon_{2}}+\frac{\eta
_{2}\varepsilon_{4}}{\varepsilon_{2}}-\eta_{4}\right) \nonumber\\
&  +\alpha\left(  102\varepsilon_{4}+2\frac{\eta_{4}\varepsilon_{4}%
}{\varepsilon_{2}}-3\eta_{6}-18\varepsilon_{4}\right) \nonumber\\
&  -16\beta^{2}\alpha\varepsilon_{8}-64\beta\alpha^{2}\varepsilon
_{10}-64\alpha^{3}\varepsilon_{12}\bigg]\nonumber\\
&  -\alpha\varepsilon_{6}\left(  2\frac{A^{\prime}}{A}+7\frac{T^{\prime}}%
{T}\right)  -\alpha n^{\prime}\partial_{n}\varepsilon_{6}\nonumber\\
&  -DT^{-2}\left[  -\frac{3}{4}\varepsilon_{0}+\frac{3}{8}\eta_{2}-\frac{1}%
{8}\frac{\varepsilon_{4}}{\varepsilon_{2}}\eta_{0}\right. \nonumber\\
&  \left.  +8\beta\alpha\left(  \varepsilon_{6}+2\frac{\varepsilon_{4}^{2}%
}{\varepsilon_{2}}\right)  +24\alpha^{2}\left(  \varepsilon_{8}+\frac
{\varepsilon_{6}\varepsilon_{4}}{\varepsilon_{2}}\right)  \right] \nonumber\\
&  -\frac{\gamma}{8}A^{2}\left(  3\mu_{2}-\frac{\mu_{0}\varepsilon_{4}%
}{\varepsilon_{2}}\right)  \bigg\}\left/  \left(  \varepsilon_{6}%
-\frac{\varepsilon_{4}^{2}}{\varepsilon_{2}}\right)  \right.  , \label{ext_a}%
\end{align}%
\begin{align}
\beta^{\prime}  &  =2\frac{T^{\prime}}{T}\beta-DT^{-2}\left(  \frac{1}{4}%
\frac{\eta_{0}}{\varepsilon_{2}}-4\beta^{2}-48\alpha^{2}\frac{\varepsilon_{6}%
}{\varepsilon_{2}}-32\alpha\beta\frac{\varepsilon_{4}}{\varepsilon_{2}}\right)
\nonumber\\
&  -\frac{\gamma}{4}A^{2}\frac{\mu_{0}}{\varepsilon_{2}}+g_{\omega}%
T^{-2}\left(  \beta\frac{\varepsilon_{0}}{\varepsilon_{2}}-\beta\frac{\eta
_{2}}{\varepsilon_{2}}+18\alpha-2\alpha\frac{\eta_{4}}{\varepsilon_{2}}\right)
\nonumber\\
&  -2\frac{\varepsilon_{4}}{\varepsilon_{2}}\left(  \alpha^{\prime}%
-4\frac{T^{\prime}}{T}\alpha\right)  , \label{ext_b}%
\end{align}
where the prime denotes a partial derivative with respect to $z$. The weighing
coefficients are given by
\begin{align}
\varepsilon_{k}\left(  n\right)   &  =\int_{-\infty}^{\infty}\tau^{k}%
p_{n}\left(  \tau\right)  \mathrm{d}\tau,\nonumber\\
\mu_{k}\left(  n\right)   &  =\int_{-\infty}^{\infty}\tau^{k}p_{n}^{2}\left(
\tau\right)  \mathrm{d}\tau,\nonumber\\
\eta_{k}\left(  n\right)   &  =\int_{-\infty}^{\infty}\tau^{k}p_{n}\left(
\tau\right)  ^{-1}\left[  \partial_{\tau}p_{n}\left(  \tau\right)  \right]
^{2}\mathrm{d}\tau. \label{coeff}%
\end{align}
To increase numerical efficiency, they are calculated only once for a
sufficiently closely spaced $n$ grid and tabulated. The evolution equations
Eqs.\thinspace(\ref{ext_n}) - (\ref{ext_b}) are also valid for $z$ dependent
coefficients in Eq.\thinspace(\ref{nse}), which is especially important for
effects like gain saturation. We note that the validity of the derived
equations is not restricted to ansatz Eq.\thinspace(\ref{pn2}); in fact, they
can be used for any such test pulse $p_{n}$ with a continuously adjustable
pulse shape parameter $n$ (and $p_{n}\left(  \tau\right)  =p_{n}\left(
-\tau\right)  $), like the q-Gaussian function \cite{boz97,nic08}. Only the
weighing coefficients $\varepsilon_{k}\left(  n\right)  $, $\mu_{k}\left(
n\right)  $ and $\eta_{k}\left(  n\right)  $ (Eq.\thinspace(\ref{coeff})) have
then to be recalculated for that specific function.

\section{Results}

To validate the ansatz Eq.\thinspace(\ref{ansatz}), the equations of motion
Eqs.\thinspace(\ref{ext_n}) - (\ref{ext_b}) are solved in different nonlinear
propagation regimes. First, the soliton regime is considered, characterized by
negative dispersion and moderate nonlinearity. Then, the self-similar
propagation through gain fibers with positive dispersion is studied. Finally,
the ansatz is employed to find the steady state solution of a
soliton-similariton fiber laser, where alternate propagation in both regimes
occurs. The equations of motion Eqs.\thinspace(\ref{ext_n}) - (\ref{ext_b})
are solved with a standard differential equation solver, allowing for an
efficient treatment of the problem. For comparison, also the results for the
simplified Gaussian ansatz Eq.\thinspace(\ref{gauss}) are shown. The
corresponding equations of motion \cite{jir06} can be obtained from
Eqs.\thinspace(\ref{ext_T}), (\ref{ext_A}) and (\ref{ext_b}) by setting $n=1$,
$\alpha=0$ and $n^{\prime}=\alpha^{\prime}=0$. The semi-analytic results are
validated against exact analytical solutions of Eq.\thinspace(\ref{nse}) or
full numerical simulations, performed with a standard symmetric split-step
propagation algorithm \cite{agr89}.

\subsection{Fundamental Soliton}

\begin{figure}[b]
\includegraphics{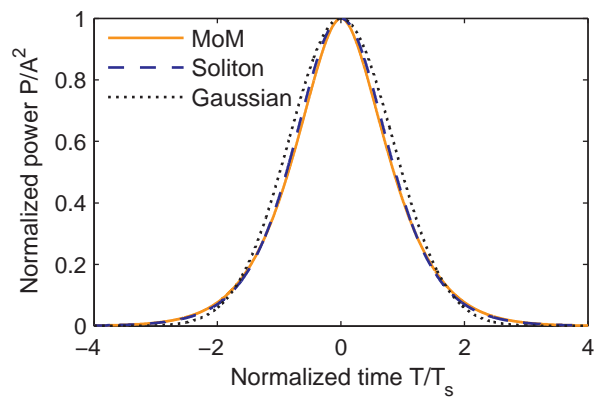}
\caption{Instantaneous power vs. time for the approximate and exact
fundamental soliton solution; for comparison, also the Gaussian approximation
is displayed.}%
\label{fig1}%
\end{figure}

For $g=g_{\omega}=r=0$, steady state solutions of Eq.\thinspace(\ref{nse})
exist. For $\gamma>0$, $D<0$ (or $\gamma<0$, $D>0$), a special solution is
given in form of the fundamental soliton, with the power $\left|  u\right|
^{2}=A^{2}\operatorname{sech}^{2}\left(  t/T_{\mathrm{s}}\right)  $, where
$T_{\mathrm{s}}=A^{-1}\left(  -2D/\gamma\right)  ^{1/2}$~\cite{agr89}. To test
the validity of our ansatz Eq.\thinspace(\ref{ansatz}), we extract the steady
state solution of the evolution equations Eqs.\thinspace(\ref{ext_n}) -
(\ref{ext_b}) with $g=g_{\omega}=r=0$, and compare it to the exact soliton
solution. By setting $\partial_{z}=0$, we obtain $\beta=\alpha=0$, $\mu
_{2}\eta_{0}+2\varepsilon_{0}\mu_{0}-\eta_{2}\mu_{0}=0$ which is fulfilled for
$n\approx0.518$, and $\mu_{0}\gamma A^{2}T^{2}=-\eta_{0}D$. The pulse energy
$E=A^{2}T\varepsilon_{0}$ can thus be written as $E=\varepsilon_{0}\left(
\eta_{0}/\mu_{0}\right)  ^{1/2}A\left(  -D/\gamma\right)  ^{1/2}$
$\approx2.79\,A\left(  -D/\gamma\right)  ^{1/2}$. The energy of the exact
solution of Eq.\thinspace(\ref{nse}), i.e., the fundamental soliton,
is\ $E_{\mathrm{s}}=2^{3/2}A\left(  -D/\gamma\right)  ^{1/2}$, thus we have
$E\approx0.99\,E_{\mathrm{s}}$. The Gaussian ansatz, Eq.\thinspace
(\ref{gauss}), is less accurate, yielding $E\approx1.05\,E_{\mathrm{s}}$. In
Fig.\thinspace\ref{fig1}, the approximate (solid line) and exact (dashed line)
solution are compared for a fixed pulse amplitude $A$. The results are
virtually indistinguishable, demonstating that the ansatz Eq.\thinspace
(\ref{ansatz}) works very well in the soliton regime. For comparison, also the
Gaussian steady state solution is displayed (dotted line). It provides a less
accurate but still reasonable fit, even though it naturally fails to reproduce
the characteristic $\mathrm{sech}^{2}$\ soliton shape. 

\subsection{\label{Comp2}Amplifier Similariton}

\begin{figure}[b]
\includegraphics{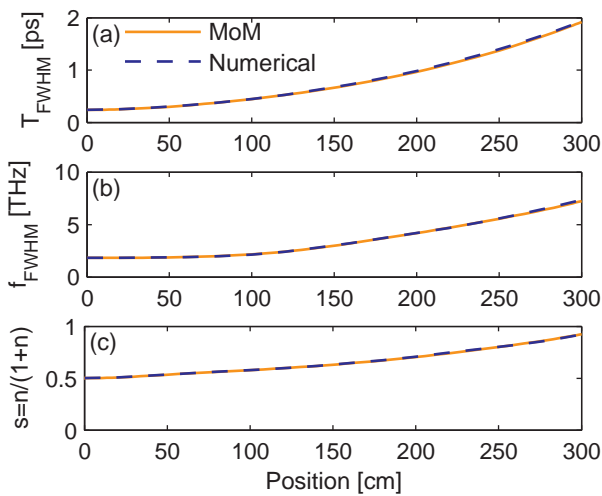}
\caption{Evolution of the pulse duration, spectral width and pulse shape as a
function of the propagation coordinate $z$, computed with the method of
moments and by solving Eq.\thinspace(\ref{nse}) numerically.}%
\label{fig2}%
\end{figure}

In order to test our ansatz in the self-similar regime, single-pass
propagation in a gain fiber with positive dispersion is studied. The
investigated setup is as described in \cite{parabolic-amp1}, with the fiber
parameter values $\gamma=5.8\times10^{-3}\,\mathrm{W}^{-1}\mathrm{m}^{-1}$,
$D=12.5\times10^{-1}\,\mathrm{ps}^{2}\mathrm{m}^{-1}$, and $g=0.95\,\mathrm{m}%
^{-1}$; furthermore, $r=g_{\omega}=0$. The initial pulse is assumed to be
Gaussian ($n=1$) with a fixed energy of $12\,\mathrm{pJ}$. First, the pulse
evolution is studied\ with ansatz Eq.\thinspace(\ref{ansatz}) and by full
numerical simulation for an initial pulse duration of $0.2\,\mathrm{ps}$.
Here, the pulse is characterized in terms of its temporal and spectral width
$T_{\mathrm{FWHM}}$ and $f_{\mathrm{FWHM}}$, respectively,\ which are the full
width at half-maximum (FWHM) values of the instantaneous power and the power
spectrum. Furthermore, $n\left(  z\right)  $ is evaluated, describing the
pulse shape of our ansatz Eq.\thinspace(\ref{ansatz}). For the numerical
pulse, the kurtosis \cite{rue06,and07} $\int\left(  t-t_{0}\right)
^{4}p\,\mathrm{d}t/\sigma^{4}$ is calculated, where $p=P/\int P\,\mathrm{d}t$
is the normalized pulse power, $t_{0}=\int tp\,\mathrm{d}t=0$ is the mean
value, and $\sigma^{2}=\int\left(  t-t_{0}\right)  ^{2}p\,\mathrm{d}t$ is the
variance; $n$ it then extracted by determining the $p_{n}$ in Eq.\thinspace
(\ref{pn2}) with the same kurtosis. In Fig.\thinspace\ref{fig2}, the evolution
of the pulse parameters is compared for the method of moments and full
numerical simulation. In Fig.\thinspace\ref{fig2}(c), $s=n/\left(  n+1\right)
$ rather than $n$ itself is plotted to restrict the range of values to
$\left[  0,1\right]  $; i.e., $s=1/2$ corresponds to a Gaussian and $s=1$ to a
parabolic pulse. In the example shown, $s$ approaches $1$, indicating that the
pulse approaches self-similar evolution. The overall agreement between
semi-analytic and numerical results is excellent, indicating that our approach
works well also in the regime of self-similar propagation. Specifically, our
ansatz Eq.\thinspace(\ref{ansatz}) fully captures the transition of the pulse
shape (see Fig.\thinspace\ref{fig2}(c)).\begin{figure}[bb]
\includegraphics{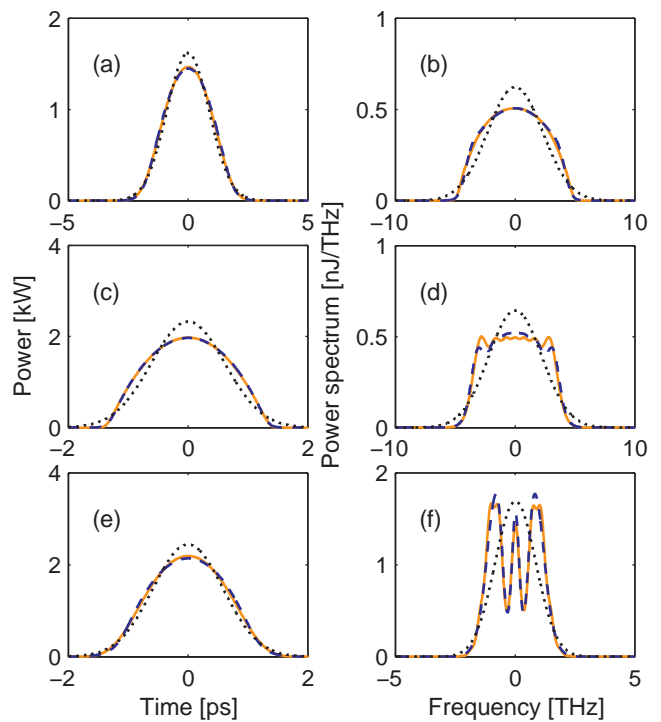}
\caption{Instantaneous power and power spectrum, as obtained with the method
of moments (solid lines), by full numerical simulations (dashed lines), and
with the simplified Gaussian ansatz (dotted lines). The initial pulse
durations are\ $0.1$, $0.2\ $and $1\,\mathrm{ps}$, respectively.}%
\label{fig3}%
\end{figure}

In Fig.\thinspace\ref{fig3}, the instantaneous power and power spectrum are
shown after a propagation distance of $3\,\mathrm{m}$ for Gaussian initial
pulse widths (FWHM) of\ $0.1\,\mathrm{ps}$ (Fig.\thinspace\ref{fig3}(a), (b))
$0.2\,\mathrm{ps}$ (Fig.\thinspace\ref{fig3}(c), (d)) and $1\,\mathrm{ps}$
(Fig.\thinspace\ref{fig3} (e), (f)), respectively. Ansatz Eq.\thinspace
(\ref{ansatz}) (solid lines) provides an excellent qualitative and
quantitative approximation, reproducing very well the exact numerical pulse
shapes and power spectra (dashed lines). The Gaussian approach (dotted lines)
shows some deviations in the pulse duration and especially the amplitude, but
overall still provides a reasonable fit in time domain, see Fig.\thinspace
\ref{fig3}(a), (c), (e). However, it naturally fails to reproduce the pulse
shapes. Especially for strongly self-similar propagation as shown in
Fig.\thinspace\ref{fig3}(c), where both our ansatz and the exact result
exhibit a distinct parabolic intensity profile, the Gaussian ansatz does not
approximate the pulse shape well. Regarding the obtained power spectra, see
Fig.\thinspace\ref{fig3}(b), (d), (f), the Gaussian ansatz completely fails to
reproduce the spectral features. The capability to faithfully reproduce
spectral characteristics is particularly important from a practical point of
view: Experimentally, optical spectra provide the most direct, immediately
available and quite informative insight into the evolution of an ultrafast
pulse among all the diagnostics at the disposal of the researcher.

\subsection{Soliton-Similariton\ Fiber Laser}

In the following, we apply our approach to self-similar propagation in a laser
cavity, where the laser field is subject to periodic boundary conditions in
steady state operation. We choose a soliton-similariton laser setup as
investigated in~\cite{SS_NP}, which is especially interesting in our context
since the pulse undergoes self-similar propagation as well as reshaping to
Gaussian and $\operatorname{sech}^{2}$ profiles in the same cavity. In our
case, the setup consists of a gain fiber, a piece of single mode fiber (SMF),
a saturable absorber (SA), a bandpass filter, and again an SMF. The pulse
evolves self-similarly in the gain fiber, and is temporally and spectrally
filtered in the SA and bandpass filter, respectively. The group velocity
dispersion (GVD) in the SMF is negative, approximately canceling the positive
GVD in the gain fiber. Several distinct nonlinear pulse shapes co-exist in the
cavity: A parabolic profile is obtained towards the end of the the gain fiber,
characteristic for self-similar evolution, then the pulse undergoes Gaussian
spectral filtering and approaches a $\operatorname{sech}^{2}$ shape in the
SMF, typical for a fundamental soliton.

The parameter values for the gain fiber (SMF) are $\gamma=9.32\times
10^{-3}\,\mathrm{W}^{-1}\mathrm{m}^{-1}$ ($1.1\times10^{-3}\,\mathrm{W}%
^{-1}\mathrm{m}^{-1}$), $D=0.03845\,\mathrm{ps}^{2}\mathrm{m}^{-1}$
($-0.0114\,\mathrm{ps}^{2}\mathrm{m}^{-1}$), $g_{0}=3.45\,\mathrm{m}^{-1}$
($0$), and $g_{\omega}=3.25\times10^{-4}\,\mathrm{ps}^{2}\mathrm{m}^{-1}$
($0$)~\cite{SS_NP}. The gain is assumed to saturate with the pulse energy $E$,
i.e., $g=g_{0}/\left(  1+E/E_{\mathrm{sat}}\right)  $, where $E_{\mathrm{sat}%
}=2.21\,\mathrm{nJ}$ is the saturation energy. The bandpass filter is modeled
by a segment of length $L$ with $g_{\omega}L=0.015\,\mathrm{ps}^{2}$,
corresponding to a spectral width of $12\,\mathrm{nm}$ (FWHM), and the pulse
power is additionally reduced by a factor of $5$ to account for the overall
linear loss of the optical cavity elements. For the SA, an unsaturated loss of
$q_{0}=0.7$ and a saturation power of $P_{\mathrm{sat}}=2.13\,\mathrm{kW}$\ is
assumed; its implementation is discussed in Appendix \ref{Ac}%
.\begin{figure}[b]
\includegraphics{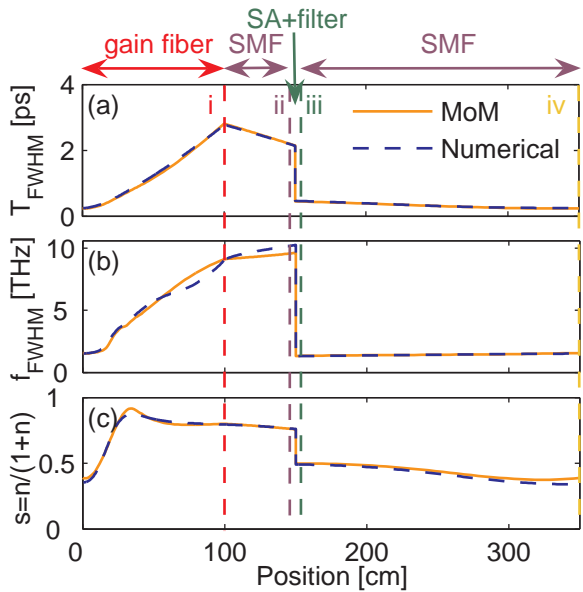}
\caption{Evolution of the pulse duration, spectral width and pulse shape in
the laser cavity, as obtained with the method of moments and by solving
Eq.\thinspace(\ref{nse}) numerically.}%
\label{fig4}%
\end{figure}\begin{figure}[bb]
\includegraphics{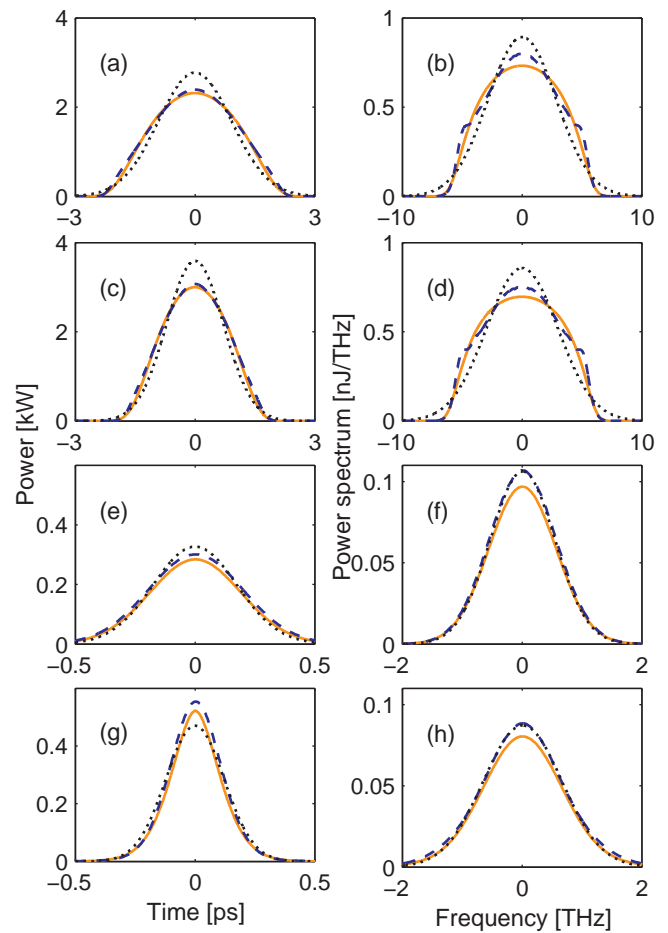}
\caption{Instantaneous power and power spectrum, as obtained with the method
of moments (solid lines), by full numerical simulations (dashed lines), and
with the simplified Gaussian ansatz (dotted lines). The results are shown at
the positions i ((a), (b)), ii ((c), (d)), iii ((e), (f)), and iv ((g), (h))
in the laser cavity, as indicated in Fig. \ref{fig4}(a).}%
\label{fig5}%
\end{figure}

In Fig.\thinspace\ref{fig4}, the MoM and full numerical results for the
evolution of characteristic pulse parameters in the cavity are compared, where
the sequence of optical elements and the fiber lengths are as indicated in
Fig.\thinspace\ref{fig4}(a). The pulse parameters are defined as described in
Section \ref{Comp2}. The overall agreement between semi-analytic and numerical
results is again excellent, compare Fig.\thinspace(\ref{fig2}). Particularly,
as can be seen in Fig.\thinspace\ref{fig4}(c), our ansatz Eq.\thinspace
(\ref{ansatz}) correctly predicts the almost parabolic pulse profile in the
gain segment, with $s=1$ for a parabolic pulse, the Gaussian shape after the
filter ($s=1/2$), as well as the $\operatorname{sech}^{2}$ shape in the SMF,
corresponding to $s\approx1/3$. In Fig.\thinspace\ref{fig5}, the instantaneous
power and power spectrum are shown after the gain fiber, before the SA, after
the bandpass filter, and after the second SMF. The overall agreement between
semi-analytic (solid lines) and numerical results (dashed lines) is very good
both in the gain fiber and the SMF. Especially, ansatz Eq.\thinspace
(\ref{ansatz}) approximates well the distinct temporal and spectral pulse
shapes in the different regimes. For comparison, also the Gaussian solution is
displayed (dotted line). It provides a reasonable fit to the temporal and
spectral width, but naturally cannot reproduce the pulse shape at all. Only
after the bandpass filter, which forces the power spectrum to assume a
Gaussian profile, the Gaussian ansatz closely matches the numerical solution
(see Fig.\thinspace\ref{fig5}(f), (h)).

\section{Conclusion}

In conclusion, we have developed a semi-analytic theory for nonlinear optical
ultrafast pulse propagation in the self-similar and other regimes, which we
employ to study the pulse dynamics in similariton amplifiers and lasers. The
key is the introduction of a model pulse with adaptive shape, which can
continuously be tweaked with a single parameter to represent pulse shapes
ranging from parabolic to Gaussian to $\mathrm{sech}^{2}$-like intensity
profiles. Thus, very different regimes of nonlinear optical propagation can be
covered. Based on the method of moments, evolution equations are derived for
the characteristic pulse parameters, specifying the pulse amplitude, duration,
profile, and linear and third order chirp. Comparison to exact analytical or
full numerical results were performed for the soliton regime as well as
similariton amplifiers and soliton-similariton lasers, showing excellent
agreement. This constitutes a semi-analytic model for the soliton-similariton
laser. A major advantage of the semi-analytic method is that the calculations
are approximately\ $100$ times faster than the full numerical simulations.
This will allow the exploration of a vast parameter range of interest to the
design of fiber and solid state similariton lasers. Furthermore, this approach
can be helpful for developing an intuitive understanding of the dynamics of
self-similar evolution in optical fiber systems by exposing the relative
importance of the various effects. Due to the versatility of our test
function, we expect it to also prove useful in other application areas in
nonlinear optics, or in completely different fields such as the description of
trapped Bose-Einstein condensates, as already exemplified in \cite{kec07}.%

\begin{acknowledgments}
C.J. acknowledges support from the German Research Foundation (DFG) within the
Emmy Noether program (JI 115/1-1) and under DFG Grant No.\,JI 115/2-1.
F.\"{O}.I. acknowledges support by the Scientific and Technological Research
Council of Turkey (T\"{U}B\.{I}TAK) Project No.\,109T350 and Project
No.\,209T058 and by the EU 7th Framework Project CROSS TRAP Grant
No.\,244068.
\end{acknowledgments}  

\appendix

\section{\label{Aa}Test Pulse}

The test pulse Eq.\thinspace(\ref{ansatz}) can be written as%
\begin{align*}
p_{n}\left(  \tau\right)   &  =\left(  1-\tau^{2}\right)  \exp\left\{
\frac{\left|  \tau\right|  ^{2n}}{n}\left[  _{2}\mathrm{F}_{1}\left(
1,n;1+n;\tau^{2}\right)  -1\right]  \right\} \\
&  =\left(  1-\tau^{2}\right)  \exp\left\{  \left|  \tau\right|  ^{2n}\left[
\Phi\left(  \tau^{2},1,n\right)  -n^{-1}\right]  \right\}  ,
\end{align*}
where $_{2}$\textrm{F}$_{1}$ is the Gauss hypergeometric function and $\Phi$
is the Lerch Phi function, defined as $\Phi\left(  z,\alpha,n\text{ }\right)
=\sum_{k\geq0}z^{k}/\left(  n+k\right)  ^{\alpha}$ for $\left|  z\right|  <1$
and analytic continuation otherwise. For $\tau^{2}=1$, where $_{2}%
\mathrm{F}_{1}$ and $\Phi$ both diverge, $p_{n}$ has to be expressed in terms
of the digamma function $\Psi\left(  z\right)  $ and Euler's constant $\gamma
$, $p_{n}\left(  \pm1\right)  =\exp\left(  -\Psi\left(  n+1\right)
-\gamma\right)  $.

These special functions are routinely implemented in many mathematical tools,
and efficient routines are available \cite{nr}. However, we found it
convenient to evaluate Eq.\thinspace(\ref{ansatz}) by a series approach, using%
\[
p_{n}\left(  \tau\right)  =\left(  1-\tau^{2}\right)  \exp\left(  \left|
\tau\right|  ^{2n}\sum_{m\geq1}\frac{\left|  \tau\right|  ^{2m}}{m+n}\right)
\]
for $\tau^{2}<1$ and%
\begin{align*}
p_{n}\left(  \tau\right)   &  =\left(  \tau^{2}-1\right)  \exp\left(  \left|
\tau\right|  ^{2n}\sum_{m\geq0}\frac{\left|  \tau\right|  ^{-2m}}{m-n}\right)
\\
&  \times\exp\left\{  \pi\frac{\cos\left(  2\pi n\right)  }{\sin\left(  \pi
n\right)  }+\pi\left[  2\cos\left(  \pi n\right)  -1\right]  \tan\left(
\frac{3}{2}\pi n\right)  \right\}
\end{align*}
for $\tau^{2}>1$ (and $n\notin\mathbb{N}$). For $n\in\mathbb{N}$, $p_{n}$\ is
directly given by Eq.\thinspace(\ref{pn}).

\section{\label{Ab}Derivation of the Equations of Motion}

The equations of motion for the pulse parameters are derived using the method
of moments \cite{MoM,tso06}. We introduce the energy $Q_{0}$ and the momentum
$P_{0}$,%
\begin{align*}
Q_{0}  &  =\int_{-\infty}^{\infty}\left|  u\right|  ^{2}\mathrm{d}t,\\
P_{0}  &  =\frac{1}{2}\int_{-\infty}^{\infty}\left(  u_{t}^{\ast}%
u-u_{t}u^{\ast}\right)  \mathrm{d}t,
\end{align*}
and higher-order generalized moments%
\begin{align*}
Q_{1}  &  =\int_{-\infty}^{\infty}t\left|  u\right|  ^{2}\mathrm{d}t,\\
Q_{\ell}  &  =\int_{-\infty}^{\infty}\left(  t-t_{0}\right)  ^{\ell}\left|
u\right|  ^{2}\mathrm{d}t,\ \ell>1\\
P_{\ell}  &  =\int_{-\infty}^{\infty}\left(  t-t_{0}\right)  ^{\ell}\left(
u_{t}u^{\ast}-u_{t}^{\ast}u\right)  \mathrm{d}t,\ \ell>0
\end{align*}
where $t_{0}$ denotes the center of gravity. Due to the symmetry properties of
the ansatz Eq.\thinspace(\ref{ansatz}), we have $Q_{\ell}=0$ for odd $\ell$
and $P_{\ell}=0$ for even $\ell$, as well as $t_{0}=0$.

Multiplying Eq.\thinspace(\ref{nse}) with $u^{\ast}$ and subtracting the
complex conjugate, we can write%
\begin{equation}
\mathrm{i}\partial_{z}\left|  u\right|  ^{2}+D\partial_{t}\left(
u\partial_{t}u^{\ast}-u^{\ast}\partial_{t}u\right)  =u^{\ast}R-uR^{\ast},
\end{equation}
with the dissipative term $R=\mathrm{i}\left(  g+g_{\omega}\partial_{t}%
^{2}+r\left|  u\right|  ^{2}\right)  u$. Multiplying with $t^{\ell}$ and
integrating over $t$ yields the equations of motion for the $Q_{\ell}$.
Furthermore, multiplying Eq.\thinspace(\ref{nse})$\ $with $u_{t}^{\ast}\ $and
subtracting $u^{\ast}$ times the temporal derivative of Eq.\thinspace
(\ref{nse}), and subsequently taking the real part of the resulting equation
yields%
\begin{align}
&  \mathrm{i}\partial_{z}\left(  u_{t}^{\ast}u-u_{t}u^{\ast}\right)
-4D\partial_{t}\left|  u_{t}\right|  ^{2}+D\partial_{t}^{3}\left|  u\right|
^{2}-\gamma\partial_{t}\left|  u\right|  ^{4}\nonumber\\
&  =2\left(  u_{t}R^{\ast}+u_{t}^{\ast}R\right)  -\partial_{t}\left(
uR^{\ast}+u^{\ast}R\right)  .
\end{align}
Multiplying with $t^{\ell}$ and integrating over $t$ yields the equations of
motion for the $P_{\ell}$. We arrive at the evolution equations%
\begin{equation}
\partial_{z}Q_{0}=\mathrm{i}\int_{-\infty}^{\infty}\left(  uR^{\ast}-u^{\ast
}R\right)  \mathrm{d}t,
\end{equation}%

\begin{equation}
\partial_{z}Q_{2}=2\mathrm{i}DP_{1}+\mathrm{i}\int_{-\infty}^{\infty}%
t^{2}\left(  uR^{\ast}-u^{\ast}R\right)  \mathrm{d}t,
\end{equation}%
\begin{equation}
\mathrm{i}\partial_{z}Q_{4}+4DP_{3}=\int_{-\infty}^{\infty}t^{4}\left(
u^{\ast}R-uR^{\ast}\right)  \mathrm{d}t,
\end{equation}%
\begin{align}
\partial_{z}P_{1}  &  =\mathrm{i}\int_{-\infty}^{\infty}\left(  -4D\left|
u_{t}\right|  ^{2}-\gamma\left|  u\right|  ^{4}\right)  \mathrm{d}t\nonumber\\
&  +2\mathrm{i}\int_{-\infty}^{\infty}t\left(  u_{t}R^{\ast}+u_{t}^{\ast
}R\right)  \mathrm{d}t+\mathrm{i}\int_{-\infty}^{\infty}\left(  uR^{\ast
}+u^{\ast}R\right)  \mathrm{d}t,
\end{align}%
\begin{align}
&  -\mathrm{i}\partial_{z}P_{3}+12D\int_{-\infty}^{\infty}t^{2}\left|
u_{t}\right|  ^{2}\mathrm{d}t-6DQ_{0}\nonumber\\
&  +3\gamma\int_{-\infty}^{\infty}t^{2}\left|  u\right|  ^{4}\mathrm{d}%
t\nonumber\\
&  =2\int_{-\infty}^{\infty}t^{3}\left(  u_{t}R^{\ast}+u_{t}^{\ast}R\right)
\mathrm{d}t+3\int_{-\infty}^{\infty}t^{2}\left(  uR^{\ast}+u^{\ast}R\right)
\mathrm{d}t.
\end{align}

Inserting Eq.\thinspace(\ref{ansatz}), we obtain
\begin{align}
&  \varepsilon_{0}\left(  2\frac{A^{\prime}}{A}+\frac{T^{\prime}}{T}\right)
+n^{\prime}\partial_{n}\varepsilon_{0}\nonumber\\
&  =2g\varepsilon_{0}+2rA^{2}\mu_{0}\nonumber\\
&  +g_{\omega}T^{-2}\left(  -\frac{1}{2}\eta_{0}-8\beta^{2}\varepsilon
_{2}-32\alpha^{2}\varepsilon_{6}-32\beta\alpha\varepsilon_{4}\right)  ,
\label{ext_Q0}%
\end{align}%

\begin{align}
&  \varepsilon_{2}\left(  2\frac{A^{\prime}}{A}+3\frac{T^{\prime}}{T}\right)
+n^{\prime}\partial_{n}\varepsilon_{2}\nonumber\\
&  =-8DT^{-2}\left(  \beta\varepsilon_{2}+2\alpha\varepsilon_{4}\right)
+2g\varepsilon_{2}+2rA^{2}\mu_{2}\nonumber\\
&  +2g_{\omega}T^{-2}\left(  -\frac{1}{4}\eta_{2}+\varepsilon_{0}-4\beta
^{2}\varepsilon_{4}-16\alpha^{2}\varepsilon_{8}-16\beta\alpha\varepsilon
_{6}\right)  , \label{ext_Q2}%
\end{align}%
\begin{align}
&  \varepsilon_{4}\left(  2\frac{A^{\prime}}{A}+5\frac{T^{\prime}}{T}\right)
+n^{\prime}\partial_{n}\varepsilon_{4}+16DT^{-2}\left(  \beta\varepsilon
_{4}+2\alpha\varepsilon_{6}\right) \nonumber\\
&  =2g\varepsilon_{4}+2rA^{2}\mu_{4}+g_{\omega}T^{-2}\Big(-\frac{1}{2}\eta
_{4}+12\varepsilon_{2}-8\beta^{2}\varepsilon_{6}\nonumber\\
&  -32\alpha^{2}\varepsilon_{10}-32\beta\alpha\varepsilon_{8}\Big),
\label{ext_Q4}%
\end{align}%
\begin{align}
&  \left(  2\frac{A^{\prime}}{A}+\frac{T^{\prime}}{T}\right)  \left(
\beta\varepsilon_{2}+2\alpha\varepsilon_{4}\right)  +\beta^{\prime}%
\varepsilon_{2}+\beta n^{\prime}\partial_{n}\varepsilon_{2}\nonumber\\
&  +2\alpha^{\prime}\varepsilon_{4}+2\alpha n^{\prime}\partial_{n}%
\varepsilon_{4}\nonumber\\
&  =-DT^{-2}\left(  \frac{1}{4}\eta_{0}+4\beta^{2}\varepsilon_{2}%
+16\beta\alpha\varepsilon_{4}+16\alpha^{2}\varepsilon_{6}\right) \nonumber\\
&  -\frac{\gamma}{4}A^{2}\mu_{0}+2g\beta\varepsilon_{2}+4g\alpha
\varepsilon_{4}+2rA^{2}\beta\mu_{2}+4rA^{2}\alpha\mu_{4}\nonumber\\
&  +g_{\omega}T^{-2}\Big(3\beta\varepsilon_{0}-\frac{3}{2}\beta\eta
_{2}+42\alpha\varepsilon_{2}-3\alpha\eta_{4}-48\beta^{2}\alpha\varepsilon
_{6}\nonumber\\
&  -96\beta\alpha^{2}\varepsilon_{8}-8\beta^{3}\varepsilon_{4}-64\alpha
^{3}\varepsilon_{10}\Big), \label{ext_P1}%
\end{align}%
\begin{align}
&  \left(  2\frac{A^{\prime}}{A}+3\frac{T^{\prime}}{T}\right)  \left(
\beta\varepsilon_{4}+2\alpha\varepsilon_{6}\right)  +\beta^{\prime}%
\varepsilon_{4}+\beta n^{\prime}\partial_{n}\varepsilon_{4}+2\alpha^{\prime
}\varepsilon_{6}\nonumber\\
&  +2\alpha n^{\prime}\partial_{n}\varepsilon_{6}+\frac{3}{4}\gamma A^{2}%
\mu_{2}+3DT^{-2}\Big(-\frac{1}{2}\varepsilon_{0}+\frac{1}{4}\eta
_{2}\nonumber\\
&  +4\beta^{2}\varepsilon_{4}+16\beta\alpha\varepsilon_{6}+16\alpha
^{2}\varepsilon_{8}\Big)\nonumber\\
&  =2g\beta\varepsilon_{4}+4g\alpha\varepsilon_{6}+2rA^{2}\beta\mu_{4}%
+4rA^{2}\alpha\mu_{6}\nonumber\\
&  +g_{\omega}T^{-2}\Big(21\beta\varepsilon_{2}-\frac{3}{2}\beta\eta
_{4}+102\alpha\varepsilon_{4}-3\alpha\eta_{6}-48\beta^{2}\alpha\varepsilon
_{8}\nonumber\\
&  -96\beta\alpha^{2}\varepsilon_{10}-8\beta^{3}\varepsilon_{6}-64\alpha
^{3}\varepsilon_{12}\Big). \label{ext_P3}%
\end{align}

Eq.\thinspace(\ref{ext_T}) is obtained after multiplying Eq.\thinspace
(\ref{ext_Q0}) by $\varepsilon_{2}/\varepsilon_{0}$ and subtracting
Eq.\thinspace(\ref{ext_Q2}); similarly, multiplying Eq.\thinspace
(\ref{ext_Q0}) by $3\varepsilon_{2}/\varepsilon_{0}$ and subtracting
Eq.\thinspace(\ref{ext_Q2}) yields Eq.\thinspace(\ref{ext_A}). Eq.\thinspace
(\ref{ext_n}) is obtained from Eq.\thinspace(\ref{ext_Q4}) by inserting
Eqs.\thinspace(\ref{ext_T}) and (\ref{ext_A}). Furthermore, we derive
Eq.\thinspace(\ref{ext_b}) by eliminating $n^{\prime}\partial_{n}%
\varepsilon_{2}$ and $n^{\prime}\partial_{n}\varepsilon_{4}$ from
Eq.\thinspace(\ref{ext_P1}), using Eqs.\thinspace(\ref{ext_Q2}) and
(\ref{ext_Q4}), respectively. Finally, Eq.\thinspace(\ref{ext_a}) is derived
from Eq.\thinspace(\ref{ext_P3}) by eliminating $\beta n^{\prime}\partial
_{n}\varepsilon_{4}$ with Eq.\thinspace(\ref{ext_Q4}) and $\beta^{\prime}$
with Eq.\thinspace(\ref{ext_b}).

\section{\label{Ac}Modeling of the Saturable Absorber}

In the Schr\"{o}dinger equation Eq.\thinspace(\ref{nse}), instantaneously
saturable gain or loss is described by the term $\left.  \partial_{z}u\right|
_{\mathrm{sat}}=r\left|  u\right|  ^{2}u$, with the solution%
\begin{equation}
u\left(  L\right)  =\frac{u_{0}}{\sqrt{1-2rL\left|  u_{0}\right|  ^{2}}}%
\end{equation}
for an initial field $u_{0}$ and a propagation length $L$. Thus, the pulse
power $P\left(  t\right)  =\left|  u\left(  t\right)  \right|  ^{2}$ is
transformed according to%
\begin{equation}
P\left(  L\right)  =\frac{P_{0}}{1-2rLP_{0}}, \label{PL}%
\end{equation}
while the phase of $u$ is not altered. For saturable absorption ($r>0$), this
approach only works in the weak field regime, i.e., $2rLP_{0}\ll1$. More
generally, a saturable absorber (SA) can be modeled by the
expression~\cite{SS_NP}%
\begin{equation}
P_{1}=P_{0}\left(  1-\frac{q_{0}}{1+P_{0}/P_{\mathrm{sat}}}\right)
=P_{0}-q_{0}P\left(  L\right)  , \label{SA}%
\end{equation}
where $q_{0}$ is the unsaturated loss, and $P_{\mathrm{sat}}$\ is the
saturation power.

In the following, we describe how to obtain the parameter values of our test
pulse Eq.\thinspace(\ref{ansatz}) after passage through an SA of the form
Eq.\thinspace(\ref{SA}). Most straightforwardly, this could be achieved by
Taylor expansion of the pulse around its center at the in- and output of the
SA and comparison of the leading terms \cite{ant07}. Here, we aim for a more
global fitting method, consistent with the MoM. First, the equations of motion
Eqs.\thinspace(\ref{ext_n}) - (\ref{ext_A}) are solved for $r=-1/\left(
2P_{\mathrm{sat}}L\right)  $ and $g=g_{\omega}=D=\gamma=0$, yielding the pulse
parameters $A\left(  L\right)  $, $T\left(  L\right)  $ and $n\left(
L\right)  $ of $P\left(  L\right)  $ in Eq.\thinspace(\ref{SA}). The
corresponding parameters $A_{1}$, $T_{1}$ and $n_{1}$ for $P_{1}$ are then
derived by computing the $0$th, $2$nd and $4$th moment of Eq.\thinspace
(\ref{SA}),%
\begin{equation}
\nu_{m}=\int_{-\infty}^{\infty}t^{m}P_{1}\mathrm{d}t=\int_{-\infty}^{\infty
}t^{m}P_{0}\mathrm{d}t-q_{0}\int_{-\infty}^{\infty}t^{m}P\left(  L\right)
\mathrm{d}t
\end{equation}
with $m=0$, $2$ and $4$, yielding%
\begin{align}
\nu_{m}  &  =A_{0}^{2}T_{0}^{m+1}\varepsilon_{m}\left(  n_{0}\right)
-q_{0}A^{2}\left(  L\right)  T^{m+1}\left(  L\right)  \varepsilon_{m}\left(
n\left(  L\right)  \right) \nonumber\\
&  =A_{1}^{2}T_{1}^{m+1}\varepsilon_{m}\left(  n_{1}\right)  ,
\end{align}
with $\varepsilon_{m}$ defined in Eq.\thinspace(\ref{coeff}). From this, we
obtain an implicit equation for $n_{1}$,%
\begin{equation}
\frac{\varepsilon_{0}\left(  n_{1}\right)  \varepsilon_{4}\left(
n_{1}\right)  }{\varepsilon_{2}^{2}\left(  n_{1}\right)  }=\frac{\nu_{0}%
\nu_{4}}{\nu_{2}^{2}},
\end{equation}
and furthermore%
\begin{align}
T_{1}  &  =\sqrt{\frac{\varepsilon_{0}\left(  n_{1}\right)  \nu_{2}%
}{\varepsilon_{2}\left(  n_{1}\right)  \nu_{0}}},\\
A_{1}  &  =\sqrt{\frac{\nu_{0}}{T_{1}\varepsilon_{0}\left(  n_{1}\right)  }}.
\end{align}
The phase $\mathrm{i}\beta\left(  t/T\right)  ^{2}+\mathrm{i}\alpha\left(
t/T\right)  ^{4}+\mathrm{i}\phi$ of our test pulse Eq.\thinspace(\ref{ansatz})
is not altered, thus we get $\beta_{1}=\beta_{0}\left(  T_{1}/T_{0}\right)
^{2}$, $\alpha_{1}=\alpha_{0}\left(  T_{1}/T_{0}\right)  ^{4}$.

\end{document}